\documentclass{mn2e}
\usepackage{times}

\usepackage{amsmath}
\usepackage{amsfonts}
\usepackage{graphicx}
\usepackage{bm}

\newcommand{\al}{\alpha}
\newcommand{\ep}{\epsilon}

\newcommand{\rh}{\rho}
\newcommand{\ph}{\phi}
\newcommand{\et}{\eta}

\newcommand{\Om}{\Omega}

\newcommand{\fr}{\frac}

\newcommand{\fot}{\frac{1}{2}}
\newcommand{\rb}{\right)}

\newcommand{\lb}{\left(}

\newcommand{\bi}{\bibitem}

\newcommand{\be}{\begin{equation}}
\newcommand{\ee}{\end{equation}}
\newcommand{\ba}{\begin{eqnarray}}
\newcommand{\ea}{\end{eqnarray}}

\newcommand{\eq}{\equiv}

\begin{document}
\journal{~}

\title[Thawing quintessence models]{Observational constraints on
  thawing quintessence models}  
\author[Timothy G. Clemson and Andrew R. Liddle]
{Timothy G. Clemson and Andrew R. Liddle\\ 
Astronomy Centre, University of Sussex, Brighton BN1 9QH, United
Kingdom}
\maketitle
\begin{abstract}
We use a dynamical systems approach to study thawing quintessence
models, using a multi-parameter extension of the exponential potential
which can approximate the form of typical thawing potentials.  We
impose observational constraints using a compilation of current data,
and forecast the tightening of constraints expected from future dark
energy surveys, as well as discussing the relation of our results to
analytical constraints already in the literature.
\end{abstract}
\begin{keywords}
cosmology: theory
\end{keywords}
%%%%%%%%%%%%%%%%%%%%%%%%%%%%%%%%%%%%%%%%%%%%%%%%%%%%%%%%%%%%%%%%%%%%%%

\section{Introduction}

Thawing quintessence models, in the terminology of Caldwell \& Linder
(2005), are those in which at early times the dark energy has a much
lower density than matter, but where the dark energy density begins to
evolve once it becomes a significant fraction of the total
(Steinhardt, Wang \& Zlatev 1999; Huey \& Lidsey 2001; Nunes \&
Copeland 2002). It is characterized by an equation of state which is
initially $w=-1$ to high accuracy, and which then `thaws' to $w>-1$ by
the present. Provided the thawing process is sufficiently slow, these
models are in agreement with current observational data.

At first sight the initial conditions for thawing quintessence appear
very unnatural, as they require the same fine-tuning of a small
initial density that one finds with the cosmological constant. This
fine-tuning could presumably be explained using the same sort of
anthropic string landscape argument often invoked for a pure
cosmological constant (Bousso \& Polchinski 2000; Susskind
2003). However the case of thawing quintessence may in fact be more
appealing, because at least for very steep potentials one can argue
that quantum fluctuations of the quintessence field acquired during
early Universe inflation may drive the field to low energy densities
\cite{ma02}. A complete cosmology from inflation through to the
present may well lead to a significant probability of thawing
quintessence behaviour, without imposing additional anthropic
constraints.

In light of this, it is useful to characterize the types of thawing
model allowed by present data, and indeed there have been several
papers recently exploring aspects of thawing quintessence models,
though these have largely ignored the issue of naturalness of initial
conditions. Inflationary models can be characterized by the slow-roll
approximation, but as this is not generally valid for quintessence
(Bludman 2004; Capone, Rubano \& Scudellaro 2006; Linder 2006; Cahn,
de Putter \& Linder 2008) some papers have sought an analogue to this
approach for dark energy \cite{sc08,du08}. Another method, employed by
Crittenden, Majerotto \& Piazza (2007), involves smoothness
requirements on the potential, while other work has developed a
classification of dark energy models by means of a calibration of
their time variation \cite{li06,li08,ca08,de08}. Analytical bounds on
thawing potentials have previously been derived by Caldwell \& Linder
(2005), Linder (2006) and Barger, Guarnaccia \& Marfatia (2006), who
also considered previous observational constraints. Expansions of the
putative quintessence potential, either using the flow equations
(Huterer \& Peiris 2007) or directly (Sahl\'en, Liddle \& Parkinson
2007) have been used to explore broad classes of models to contrast
the thawing and non-thawing regimes.

In this paper we analyze thawing models from a somewhat different
perspective. We look at a multi-parameter potential family $V(\phi)$,
which includes the exponential potential as a special case, and which
has enough freedom that the family can represent arbitrary values of
$V$, $dV/d\phi$, and $d^2V/d\phi^2$ at the initial field
value. Reminiscent of the slow-roll approximation to inflationary
observables, which depend only on the potential and its first few
derivatives, we aim to capture with this potential the full spectrum
of thawing quintessence phenomenology. This is then confronted with
present observational data, and the capabilities of future data
assessed, before a consideration of the analytical constraints derived
in previous analyses of thawing behaviour.

\section{Thawing quintessence}

\subsection{Quintessence dynamics}

We assume a minimally-coupled scalar field $\phi$ acts as
quintessence.  The pressure/density relation is $w=p/\rho$ with 
\be
p=\fr{\dot{\ph^2}}{2}-V(\ph), \quad
\rh=\fr{\dot{\ph^2}}{2}+V(\ph),\label{ed}\ee where dots denote
derivatives with respect to time and $V(\ph)$ is the field's
self-interaction potential. The variation in $\phi$ obeys the
Klein--Gordon equation \be\ddot{\ph}+3H\dot{\ph}+V'=0,\label{kg}\ee
where the prime indicates a derivative with respect to $\ph$. This
shows that the field rolls down the slope of its potential $V(\ph)$
and that its motion is damped by the Hubble parameter which, in units
with $8\pi G=1$, may be written \be
H=\fr{\dot{a}}{a}=\sqrt{\fr{\rh_t}{3}},\label{hb}\ee where $a$ is the
scale factor of the Universe and $\rh_t$ is the total density. We
assume a flat matter-dominated Universe throughout.

The following dynamical systems approach, equations~(\ref{xd}) through
(\ref{eos}), was developed by Copeland, Liddle \& Wands (1998), de la
Macorra \& Piccinelli (2000), Ng, Nunes \& Rosati (2001), and Scherrer
\& Sen (2008). Equations
(\ref{kg}) and (\ref{hb}) may be rewritten in terms of three new
variables; $x$, $y$ and $\lambda$, defined as 
\be
x\eq\fr{\dot{\ph}}{\sqrt{6}H},\quad y\eq\sqrt{\fr{V(\ph)}{3H^2}},
\quad \lambda\eq -\fr{V'}{V}.
\label{xd} 
\ee
For a Universe containing only matter and a scalar field, the system
may be written as \cite{ng01}
\ba
\fr{dx}{dN}&=&-3x+\lambda\sqrt{\fr{3}{2}}y^2+\fr{3}{2}x(1+x^2-y^2), \\
\fr{dy}{dN}&=&-\lambda\sqrt{\fr{3}{2}}xy+\fr{3}{2}y(1+x^2-y^2),
\\\fr{d\lambda}{dN}&=&-\sqrt{6}\lambda^2(\Gamma-1)x,\label{lp}\ea
where $N$ is the logarithm of the scale factor, $N\eq\ln a$, and
$\Gamma$ is defined as \cite{st99}   
\be\Gamma\eq\fr{VV''}{(V')^2}\label{gd}.\ee 
For typical potentials, including the one we introduce below, $\Gamma$
is a function of $\lambda$ (though only expressible as such if
$\lambda(\phi)$ has an analytic inversion) and hence the system is an
autonomous one (see e.g.\ Fang et al.~2008), though this property is not
needed in our analysis. 

In a Universe with flat geometry, the density parameter of the scalar
field becomes 
\be\Om_\phi=x^2+y^2,\label{dp}\ee 
and its effective equation of state is 
\be
\gamma\eq1+w=\fr{2x^2}{x^2+y^2}.\label{eos}\ee

\subsection{Generalized thawing potential}

The initial value of the scalar field is assumed to be fixed at some
early time in the matter-dominated era, and can be set to zero without
any loss of generality, ie. $\ph_{\rm i}=0$. Having made this choice,
we can then explore a wide range of thawing models by taking the
quintessence potential to be of the form
\be 
\label{genpot}
V(\ph)=V_{\rm i} \exp(-c\ph)[1+\al\ph], 
\ee 
where $V_{\rm i}$ is the initial value of the potential, and $c$ and
$\al$ are parameters. This form is useful because of two properties.
Firstly, it reduces to the well-known exponential potential in the
case $\alpha = 0$; this property motivates our choice in favour of a
simple Taylor expansion around the origin, as often used in direct
reconstruction of the quintessence potential (e.g.\ Sahl\'en, Liddle
\& Parkinson 2005). Secondly, in the vicinity of the origin it can
approximate arbitrary potentials up to their second derivative.  In
the limit $\alpha \rightarrow c$ it can also approximate the hilltop
quintessence models recently studied by Dutta \& Scherrer (2008).

Our results will therefore have two separate interpretations. The
first is to consider equation (\ref{genpot}) to be an exact potential
(somewhat similar to the Albrecht--Skordis (2000) potential), valid
for all $\phi$, enabling generalization of results from the
exponential case. The second is to consider the results to be valid
for {\em all} thawing potentials, provided the variation in $\phi$ is
small enough that our potential is a good local approximation to an
arbitrary potential.  We will take the convergence condition for this
potential to be $|\al\ph | \leq 0.1$. We also restrict the parameters
to $c\geq \alpha$ so that initially $V' \leq 0$ and hence the field
value increases with time.

\begin{figure}
	\centering
		\includegraphics[width=0.9\columnwidth]{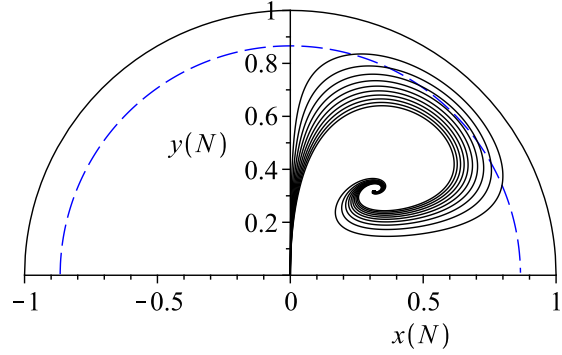}
	\caption{The evolution for $\lambda_{\rm i}=0.1$ (outer curve)
	to $1$ (inner curve) in steps of $0.1$ for $c=4$. The dashed
	semicircle represents the present-day dark energy density
	parameter $\Om_\phi\approx0.75$. Trajectories which reach this
	may correspond to the evolution of the Universe. The origin is
	the initial point at some early time during matter domination,
	and the unit semicircle represents domination by the scalar
	field.}
	\label{xy}
\end{figure}

Putting this potential into equation (\ref{lp}) leads to
\be\fr{d\lambda}{dN}=\sqrt{6}(\lambda-c)^2 x , \label{lambda}\ee 
and the initial value of $\lambda$ is given by 
\be\lambda_{\rm i}=c-\al.\ee 
We now allow the system to evolve out of an initial perturbation from
$\ph_{\rm i}=0$, exploring the parameter space by plotting the
evolution of the system over time for a range of values of $\al$ and
$c$ (e.g.\ see Fig.~\ref{xy}). Equation (\ref{dp}) implies that
$0\leq x^2+y^2\leq 1$ and so the resulting trajectories are confined
to within the unit circle. From the definitions of $x$ and $y$,
equation (\ref{xd}), it can also be seen that making $y$ negative and
reversing the temporal coordinate (and thus $\dot{\ph}\rightarrow
-\dot{\ph}$), results in symmetrical behaviour below the axis to that
above. Only the upper half-disc therefore need be considered for a
full picture of possible trajectories. Furthermore, the lower
half-disc corresponds to a contracting Universe, which is ruled out by
observations as far as the past and present Universe is concerned.

Our initial condition keeps the initial scalar field velocity at zero,
which is not quite appropriate. Cahn et al.~(2008) carried out an
analysis of the early-time behaviour of quintessence fields to compute
the leading-order early-time behaviour of the velocity. In our case
this translates as $x_{\rm i} \propto y_{\rm i}^2$, meaning that the
initial velocity is indeed highly suppressed. We confirmed numerically
that the difference to the trajectories is completely negligible.

\section{Observational constraints}

\subsection{Current constraints}

\label{s:curr}

The present-day dark energy density parameter has an observed value of
$\Om_\phi\approx0.75$ \cite{ko08} and using equation (\ref{dp}) we can
determine which trajectories reach this value and as such may be
representative of the real world (e.g.\ Fig.~\ref{xy}).

The set of possible models is better constrained, however, by
considering their representation in the $w_0$--$w_a$ plane. Thawing
dark energy is well parameterized by the equation \cite{ch01,li03} 
\be
w(a)=w_0+w_a(1-a),\label{wa}
\ee 
where $w_0$ represents the present-day
value of $w$, and $w_a$ determines the change in $w$ with the scale
factor $a$. Equation (\ref{eos}) can now be used to derive expressions
for $w_0$ and $w_a$ in terms of the variables of the autonomous
system. Since $a = 1$ at the present day we have 
\be
w_0=\fr{x_0^2-y_0^2}{x_0^2+y_0^2},
\ee 
where $x_0$ and $y_0$ are the present-day values of $x$ and $y$. The
observational constraint that $w$ is close to $-1$ therefore requires
viable trajectories to remain close to the $y$-axis at the time they
reach the semicircle indicating the correct dark energy density.

\begin{figure}
	\centering
		\includegraphics[width=0.9\columnwidth]{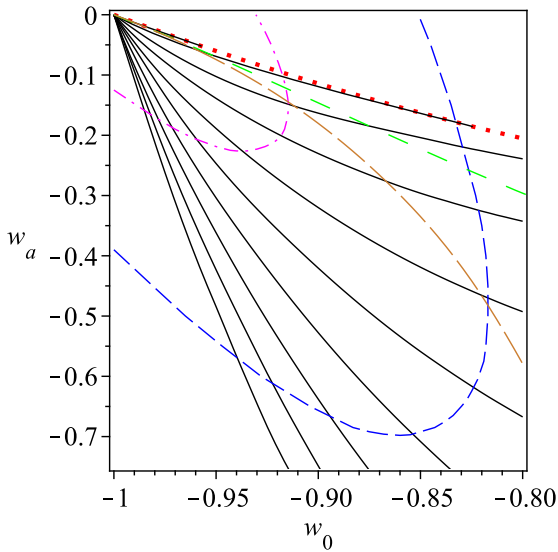}
	\caption{Families of potentials which reach $\Om_\ph=0.75$, and
	have positive values of the parameter $\alpha$, for
	$c=0.5$ (top solid line) to $5$ (lowest solid line) in steps of
	$0.5$. The short-close-dashed line delineates the range of values 
	allowed	by recent observational data and the dotted line
	represents the limit where $\al=0$. The dot-dash line 
	indicates the ten-year observational prospects discussed
	in Section~\ref{ss:future}, while the spaced-dashed and the 
	long-dashed curves represent potentials with $\Gamma_{\rm i}=0$ and
	$\Gamma_0=0$ respectively.} 
	\label{paw0wa}
\end{figure}

The time dependence of the equation of state is also observationally
constrained, though less strongly.  Combining equations (\ref{eos})
and (\ref{wa}) leads to the expression for $w_a$ 
\be
w_a=-\fr{4x_0y_0}{(x_0^2+y_0^2)^2}\lb-3x_0y_0 +
\lambda_0\sqrt{\fr{3}{2}}y_0^3  +
\lambda_0\sqrt{\fr{3}{2}}x_0^2y_0\rb,
\ee 
where $\lambda_0$ is the present-day value of $\lambda$.

\begin{figure}
	\centering
		\includegraphics[width=0.9\columnwidth]{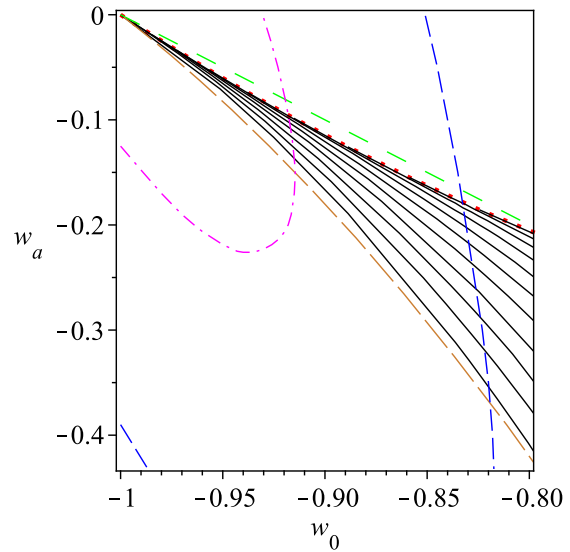}
	\caption{As Fig.~\ref{paw0wa}, but for potentials with
	negative values of the parameter $\alpha$ and
	$c=0$ (lowest solid curve) to $1$ (top solid line) in steps of
	$0.1$. The spaced-dashed line is the $-w_a\geq 1+w_0$ thawing
	limit, while the long-dashed curve represents the curve where 
	$X=3/4$ (see Section~\ref{Analytical Constraints}).} 
	\label{naw0wa}
\end{figure}

The coordinates at which trajectories reach the present dark energy
density can now be used in these two equations to plot the predictions
from families of potentials in the $w_0$--$w_a$ plane, shown in
Figs.~\ref{paw0wa} and \ref{naw0wa}. These values are compared to the
observationally-allowed region (short-close-dashed curve) obtained by
Komatsu et al.~(2008) from a combination of baryon acoustic
oscillation, type 1A supernovae and WMAP5 data at a 95\% confidence
limit, and would only be marginally improved by the inclusion of
big-bang nucleosynthesis constraints.

One might further worry whether the approximation of constant $w_a$ is
accurate enough, and indeed Dutta \& Scherrer (2008) indicated that in
some parameter regions it will not be. We quantify this below by
delineating the regions of parameter space in which $w_a$ has varied
by more than 25\% from redshift one to the present.

Each of the solid curves in Figs.~\ref{paw0wa} and \ref{naw0wa}
represents a family of potentials with the same value of the parameter
$c$ in the exponent, but different values of the linear parameter
$\alpha$ and therefore $\lambda_{\rm i}$ (remember $\lambda_{\rm
i}=c-\alpha$). They are {\em not} the evolutionary tracks of
particular potentials, but rather indicate the points in the
$w_0$--$w_a$ plane which different potentials have reached at the
present day. The largest values of $\alpha$ are at the left-hand end
of the curves in Fig.~\ref{paw0wa}, where they all share a common
origin at the point $[-1,0]$, corresponding to the cosmological
constant. Potentials at this point have $c=\alpha$ which means that
$\lambda_{\rm i}=0$ and so they are flat, while nearby potentials have
negative curvature, ie. $\Gamma<0$, and correspond to the case of
hilltop quintessence as described by Dutta \& Scherrer (2008). Moving
to the right along the curves the potentials have decreasing values of
$\alpha$ and hence increasing $\lambda_{\rm i}$, ie. increasingly
negative initial slope.

Useful orientation into the behaviour of the family of models comes
from studying the time dependence of $\Gamma$. We see from equation
(\ref{lambda}) that $\lambda$ is monotonically increasing with time,
and hence $\Gamma$ too is an increasing function whose initial value
may be positive or negative (we keep $c>0$ throughout, corresponding
to $\phi$ increasing with time). The sign of $\Gamma$ is the same as
of $V''$, and hence determines whether the potential is steepening or
become more shallow with time, with $\Gamma=0$ corresponding to a
point of inflection in the potential. We can then classify the models
as follows. If the initial value of $\Gamma$, $\Gamma_{\rm i}$, is
already positive, then $\Gamma$ stays positive forever and the
potential is becoming shallower with time (the exponential case is the
archetypal example). If the {\em present} value of $\Gamma$,
$\Gamma_0$, is negative, then $\Gamma$ has been negative throughout
the past evolution of the Universe, corresponding to a steepening
potential as in hilltop quintessence. If neither of these conditions
is satisfied, then $\Gamma$ changes from negative to positive during
the evolution, with the potential initially steepening and later
becoming more shallow.

In the figures, the curvature of the potentials increases to the right
and eventually potentials with $\Gamma_0=0$ (long-dashed) and
subsequently $\Gamma_{\rm i}=0$ (spaced-dashed line) are reached,
corresponding to points of inflection in the potential curve.  After
the points of inflection the slope of the potentials increases along
the curves, which then all head towards the exponential limit as
$\alpha$ approaches zero. They eventually graze the exponential curve
where $c=\lambda=\textrm{constant}$ and the curvature reaches its
maximum value of $\Gamma_{\rm i}=1$. Higher values of $c$ than those
plotted in Fig.~\ref{paw0wa} are possible, but only as long as
$c\approx\alpha$. For example, the trajectory of $c=20$ in the
$x$--$y$ plane only reaches the dark energy density parameter curve
when $c-\alpha\leq10^{-10}$.

\begin{figure}
	\centering
		\includegraphics[width=0.9\columnwidth]{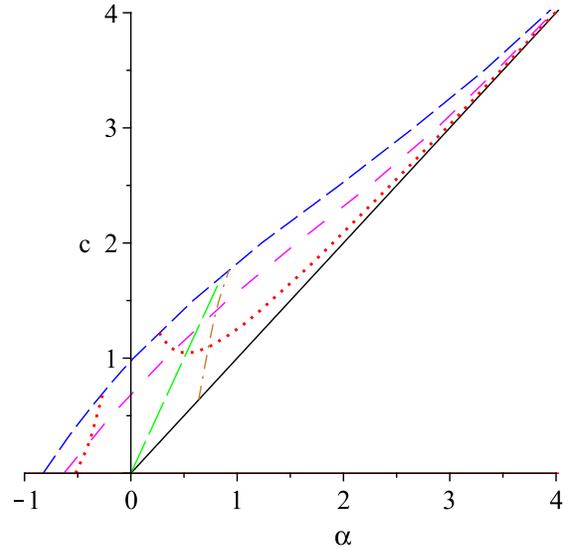}
	\caption{The allowed potentials in the $c$--$\al$ parameter
	space. The region beneath the short-dashed line is allowed by
	the observational constraints, while the spaced-dash line
	below it gives an approximation to the ten-year observational
	prospects discussed in Section~\ref{ss:future}. The region
	between the dotted lines contains potentials for which
	convergence remains good ($|\al\ph|<0.1$) up until the present
	day and the solid line represents the physical limit for which
	$\lambda$ is positive (i.e.\ $dV/d\phi$ remains
	negative). Models to the left of the dot-dashed line have
	$w_a$ constant to a good approximation; our observational
	treatment is less reliable for those to the right of this
	line. Finally, the long-dashed line is where $\Gamma_{\rm
	i}=0$, which divides the parameter space into shallowing
	potentials on the left and those which undergo a period of
	steepening on the right.}
	\label{avsc}
\end{figure}

The curves in Fig.~\ref{naw0wa} represent similar families of
potentials to those in Fig.~\ref{paw0wa} but with negative values of
the parameter $\alpha$. These curves begin from the exponential limit
at their left-hand end and have increasingly negative values of
$\alpha$ to the right. This means that they have increasing
$\lambda_{\rm i}$, ie. the initial slope of the potentials becomes
ever steeper to the right. Here the curves once again diverge from the
exponential case as the magnitude of $\alpha$ increases. This is a
consequence of their increasing linearity, which causes the curves to
asymptotically approach the family of linear potentials with
$c=\Gamma=0$, as eventually $\lambda_{\rm i}\approx c$. Eventually
each family of curves comes into conflict with observations, which
therefore give a lower limit on $\alpha$ for each value of $c$.  The
region of potentials in the $c$--$\al$ parameter space consistent with
the observational constraints is shown in Fig.~\ref{avsc}.

The range of potentials for which convergence remains good up until
the present day can be found by rewriting the convergence condition as
\be 0.9<\fr{\al}{c-\lambda_0}<1.1,\ee and is also shown in
Fig.~\ref{avsc}.  Potentials with $c<0$ are not considered here, since
they simply correspond to those for which the field rolls in the
opposite direction and so add nothing new to the analysis. Figure 4
also shows the region (essentially $\alpha \ge 1$) where the
assumption of constant $w_a$ used in our observational comparison is
breaking down; the observational results should be considered less
robust in this region.

Admitting potentials with $\al<0$ allows $V(\ph)$ to become negative
in the future, which may prompt future recollapse of the Universe
(Kallosh et al.~2003). Such potentials are included in our analysis,
though they may be considered as being approximate representations of
the behaviour of arbitrary potentials from early times to the present
day, without necessarily extrapolating into the negative energy
density regime.

\begin{figure}
	\centering
		\includegraphics[width=0.74\columnwidth]{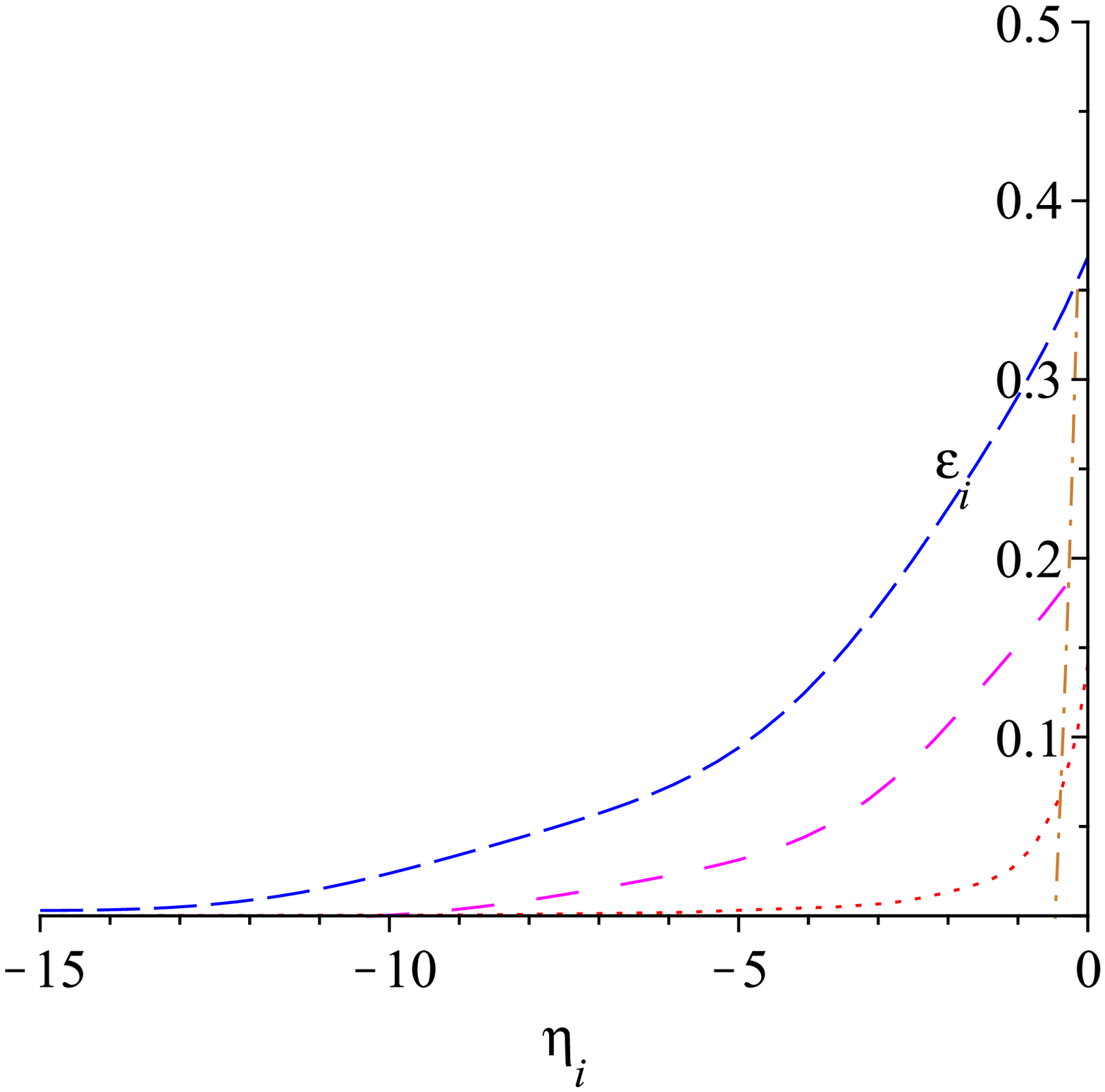} \\
		\includegraphics[width=0.74\columnwidth]{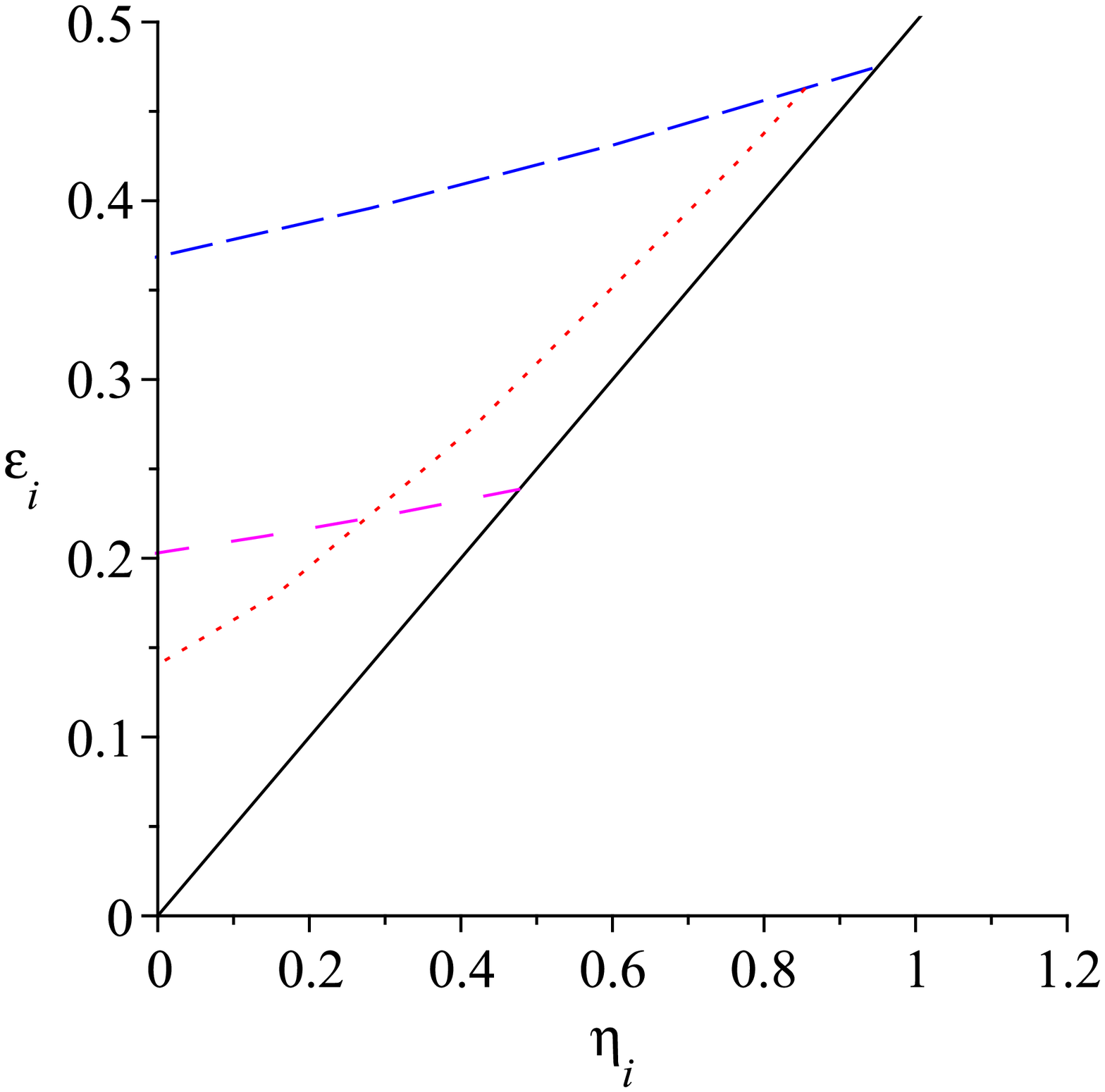}\\
		\includegraphics[width=0.74\columnwidth]{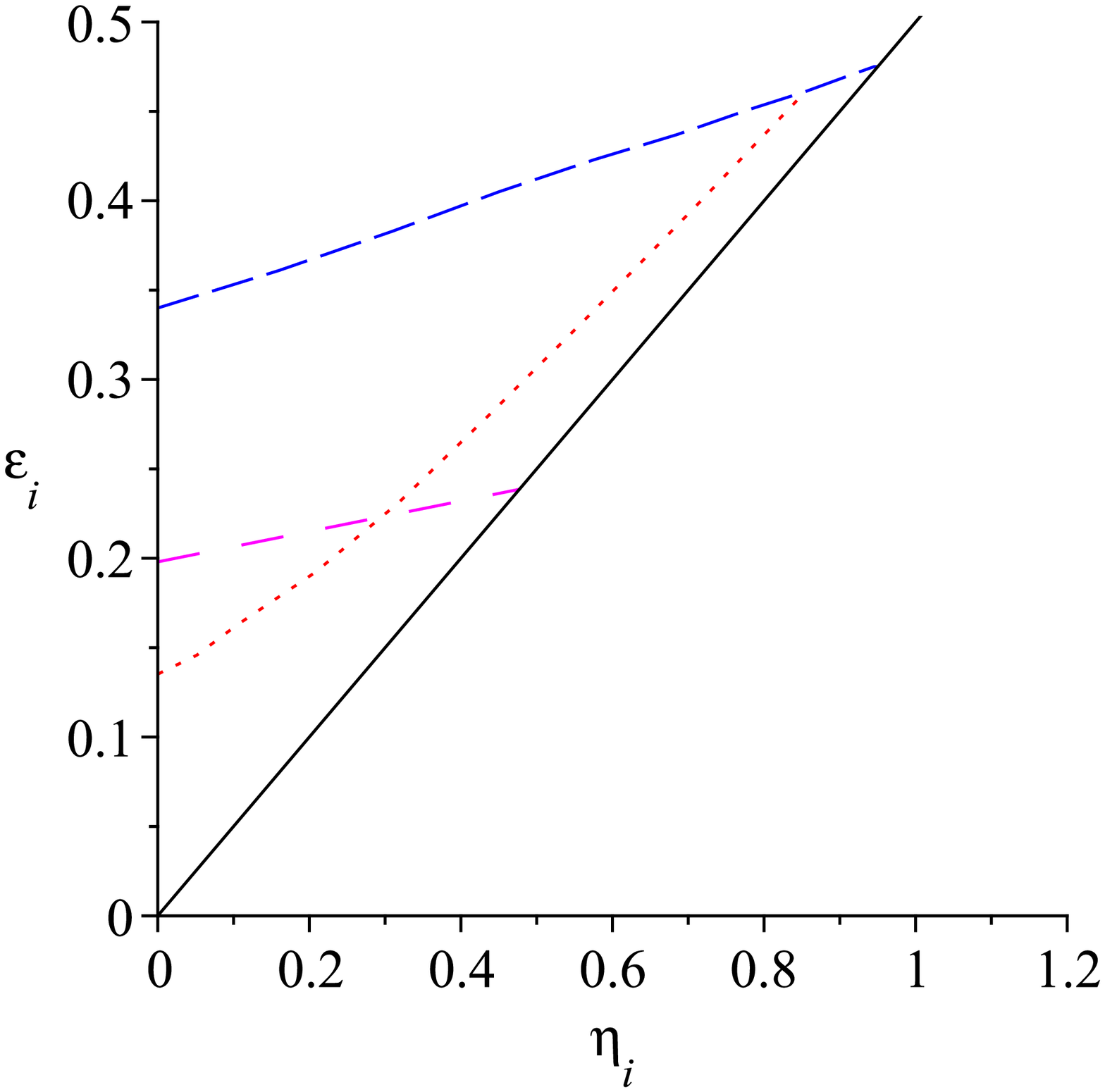}
	\caption{The allowed region of potentials in the $\et$--$\ep$
	plane.  Positive $\al$ with negative $\Gamma_i$ is shown on at
	the top, positive $\al$ with positive $\Gamma_i$ is shown in the
	middle, and negative $\al$ (all of which have $\Gamma_i>0$) are
	at the bottom. The lower two figures correspond to
	potentials which have been shallowing throughout their
	history.  The region below the dashed line is consistent with
	present observations and the space-dash line gives an
	approximation to the ten-year observational prospects
	discussed in Section~\ref{ss:future}. Potentials below the
	dotted line all have $|\al\ph|<0.1$ up until the present day,
	and so converge well. The $\ep$ axis is where $\Gamma_i=0$ and
	the $\et$ axis is where $\al=c$, while the solid line in the
	lower two figures represents the limit where $\al=0$. In the
	top figure, only models to the right of the dot-dash line have
	$w_a$ constant to a good approximation (this condition is
	always satisfied in the lower two graphs).}
	\label{evse}
\end{figure}

The representation of our results in terms of the more familiar
slow-roll parameters from inflation, $\ep$ and $\et$, is easily found
from their definitions (in natural units)
\ba\ep_{\rm i}&\eq&\fot\lb\fr{V_{\rm i}'}{V_{\rm
    i}}\rb^2=\fot\lambda_{\rm i}^2, \\
\et_{\rm i}&\eq&\fr{V_{\rm i}''}{V_{\rm i}}=2c\lambda_{\rm i}-c^2,\ea
and is shown in Fig.~\ref{evse}. As would be expected from the results
of Scherrer \& Sen (2008) and Dutta \& Scherrer (2008), the assumption
of constant $w_a$ fails for large negative $\eta$ (the top panel),
meaning our observational comparison method is not robust, but is fine
otherwise. 

\subsection{Future prospects}

\label{ss:future}

As well as the present-day observational constraints, we can use
predictions of future bounds to give a sense of how the range of
possible potentials may be further restricted. Each graph (except
Fig.~\ref{ancons}) displays an approximation to the ten-year
observational prospects (dot-dashed curve) under the assumption that
$\Lambda$CDM is the true model. This was achieved by approximating the
predicted future $w_0$--$w_a$ limits from the literature and then
resizing the present observational constraints curve to fit them,
thereby assuming similarly distributed future constraints. These
constraints roughly correspond to predictions for both forthcoming
Stage 3 programmes, such as the Dark Energy Survey, at 68\%
c.l. \cite{des} and future Stage 4 projects, such as JDEM, at 95\%
c.l. \cite{detf}. Figs. \ref{paw0wa} and \ref{naw0wa} show that a
significant limiting of the allowed region in the $w_0$--$w_a$ plane
is possible over the next ten years, but Fig. \ref{avsc} shows that
this does not translate into a correspondingly large reduction in the
$c$--$\al$ parameter space used in our model. There is however a
slightly greater reduction in the size of the allowed region in the
$\et$--$\ep$ plane, as shown in Fig. \ref{evse}.

\section{Analytical Constraints}
\label{Analytical Constraints}

\begin{figure}
	\centering
		\includegraphics[width=0.9\columnwidth]{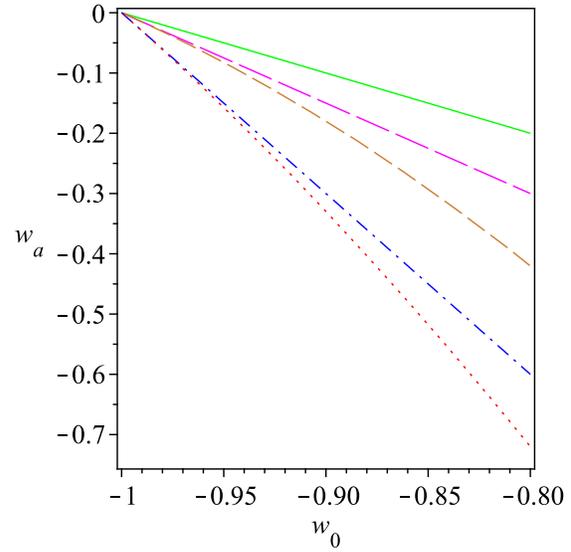}
	\caption{The solid line is the upper thawing bound of
	$-w_a\geq1+w_0$, the long-dashed line is the Barger et
	al.~(2006) limit and the short-dashed curve is where
	$X=3/4$. The dot-dashed line is the original Caldwell \&
	Linder (2005) lower thawing bound and the dotted line is the
	curve where $X=3/2$.}  
	\label{ancons}
\end{figure}

A number of different analytical constraints have been placed on
thawing models and the present situation could perhaps bear some
clarification. Fig.~\ref{ancons} shows constraints in the $w_0$--$w_a$
plane from previous analyses of thawing dynamics. The upper limit
holds in all cases, as only below this line is the requirement $w >
-1$ fulfilled, ie. $-w_a\geq1+w_0$ (solid line). Also, from the
original definition of thawing behaviour by Caldwell \& Linder (2005),
comes the lowest of the linear boundaries, $-w_a \leq 3(1+w_0)$
(dot-dashed line). This comes from the requirement that
$\ddot{\phi}t<\dot{\phi}$ for potentials where $\ddot{\phi}$ is
decreasing for all time. The slope of such potentials is monotonically
increasing (becoming less negative) as the field asymptotically
approaches the minimum of its potential curve.

Linder (2006) defined two useful ratios of the Klein--Gordon terms and
the relationship between them 
\be 
X\equiv\frac{\ddot{\phi}}{H\dot{\phi}}\,;
\quad Y \equiv \frac{\ddot{\phi}}{V'}\,;
\quad X = -3\frac{Y}{1+Y} \,.
\ee
The Klein--Gordon equation can be used along with these definitions to
obtain an expression for $X$ in terms of $x,y$ and $\lambda$ 
\be
X=3\lb\frac{\lambda y^2}{\sqrt{6}\,x}-1\rb .
\ee 
This allows us to plot given values of $X$ using the generalized
potential. The aforementioned lower bound of $-w_a \leq 3(1+w_0)$ was
shown by Linder (2006) to be an approximation to the more precise
limit of $X=3/2$ (dotted curve), although this curve does still
contain the implicit approximation that $H\approx 2/3t$. Barger et
al.~(2006) applied the original Caldwell \& Linder (2005) bound at a
redshift of 1, allowing for a significant tightening of the
constraints on potentials with monotonically increasing slope and
showed that such potentials satisfy $-w_a>(3/2)(1+w_0)$ (long-dashed
line). This is again a linear approximation and we plot this bound as
the more accurate curve where $X=3/4$ (short-dashed curve).

Potentials which steepen as the field rolls down may still satisfy the
basic requirement for thawing that $w$ begins very close to $-1$
before increasing at later times, despite permitting a more complex
thawing behaviour. In principle their range extends all the way to the
$w_0=-1$ line, which represents the divide from the present-day
phantom regime but, as mentioned in Section~\ref{s:curr}, in our model
such potentials need to be increasingly finely tuned the closer they
are to $w_0=-1$. Linder (2006) mentioned that potentials which steepen
as the field rolls down should lie below the bound of
$\ddot{\phi}t<\dot{\phi}$.  We find that this requirement however,
whilst sufficient to ensure that potentials steepen as the field rolls
down, is not a {\em necessary} condition. We find that potentials
which do undergo a period of steepening slope since the onset of
matter domination all lie beneath the curve where $\Gamma_{\rm i}=0$
(spaced-dashed line) in Fig.~\ref{paw0wa}. This represents the limit
below which the potentials have all reached the point of inflection in
the generalized potential curve by the initial early time during
matter domination. We find that this {\em coincidentally} corresponds
to a very good accuracy to the aforementioned Barger et al.~(2006)
linear bound, which may therefore be used as an upper limit for
potentials which have not had monotonically-increasing slope
throughout matter domination.

The long-dashed curve in Fig.~\ref{paw0wa} shows potentials for which
$\Gamma_0=0$, ie. they reach their point of inflection at the present
day, and as such have been steepening throughout matter domination.
Potentials between this curve and the $\Gamma_{\rm i}=0$ line are
therefore those which have crossed their point of inflection at some
time during the matter-dominated epoch, and so the distance between
these two curves is related to the age of the Universe. It should also
be noted that above the $X=3/4$ curve (long-dashed in
Fig.~\ref{naw0wa}) there are always two potentials at each point in
the $w_0$--$w_a$ plane, and above the $\Gamma_{\rm i}=0$ curve
(spaced-dashed in Fig.~\ref{paw0wa}) there are always two potentials
with monotonically increasing slope at each point.

Potentials above the $\Gamma_{\rm i}=0$ curve in Fig.~\ref{paw0wa} and
those with negative $\alpha$ (ie. all those in Fig.~\ref{naw0wa}),
have all had positive curvature ($\Gamma_{\rm i}>0$), and as such
increasingly shallower slope, since the initial time. These are the
potentials to which the Caldwell \& Linder (2005), Linder (2006), and
Barger et al.~(2006) limits all apply. We find that the lower limit on
such potentials is the case where $c=0$, i.e.\ the family of linear
potentials. Since linear potentials lie approximately along the curve
of $X=3/4$ we find that this could therefore be used as a lower bound
on potentials with monotonically increasing slope to which the
previous limits have been applied, as shown in Fig.~\ref{naw0wa}
(long-dashed curve). Also shown in Fig.~\ref{naw0wa} is our finding
that the exponential potentials (large-dotted curve) represent an
upper limit and may still be best approximated by the original
Caldwell \& Linder (2005) upper bound (spaced-dashed line).

\section{Conclusions}

The form of the scalar field self-interaction potential studied here
allows observational constraints to be placed on thawing quintessence
models, encompassing a wide range of thawing behaviour. A pure
exponential potential for example, which has the greatest
observationally-allowed slope ($\lambda_{\rm i}\approx 1$), is
currently restricted to having $c$ less than about $1$, and upcoming
observations could reduce this by up to a third. Despite applying
strict convergence conditions, we expect the chosen potential to be
fairly reliable within the whole of the observationally-allowed
region. This is especially true when considering the ten-year
observational prospects, for which the value of $|\al\ph|$ peaks at
around 0.3.

The analytical constraints on thawing models in the $w_0$--$w_a$ plane
are tightest for potentials which have had monotonically increasing
slope throughout their evolution history. Their range may be
reasonably well approximated using limits from previous analyses, but
within these bounds there are always two potentials at any given point
in the $w_0$--$w_a$ plane. Potentials which have undergone a period of
decreasing slope have an upper bound which is lower than that for
purely shallowing potentials, and which is coincidentally well
approximated by an existing limit, but only fine-tuning arguments
restrict these potentials otherwise. Future data will require new
analytical constraints to be fitted to potentials in the thawing
region for the sake of accuracy, but for now existing limits properly
applied offer a fairly accurate representation of the true range of
viable models.

\section*{Acknowledgments}

T.G.C.\ was supported by the Royal Astronomical Society and A.R.L.\ by
STFC.  We thank the referee, Bob Scherrer, for raising important points
in his report.

%%%%%%%%%%%%%%%%%%%%%%%%%%%%%%%%%%%%%%%%%%%%%%%%%%%%%%%%%%%%

\bsp

\end{document}